\documentclass[pre,twocolumn,amssymb,showpacs]{revtex4}

\usepackage{graphicx}
\usepackage{color}
\usepackage{epsfig}
\usepackage{latexsym}
\usepackage{bm}
\usepackage{ulem}

\usepackage{color}

\begin{document}

\renewcommand{\ni}{{\noindent}}
\newcommand{\dprime}{{\prime\prime}}
\newcommand{\be}{\begin{equation}}
\newcommand{\ee}{\end{equation}}
\newcommand{\bea}{\begin{eqnarray}}
\newcommand{\eea}{\end{eqnarray}}
\newcommand{\nn}{\nonumber}
\newcommand{\bk}{{\bf k}}
\newcommand{\bQ}{{\bf Q}}
\newcommand{\q}{{\bf q}}
\newcommand{\s}{{\bf s}}
\newcommand{\bN}{{\bf \nabla}}
\newcommand{\bA}{{\bf A}}
\newcommand{\bE}{{\bf E}}
\newcommand{\bj}{{\bf j}}
\newcommand{\bJ}{{\bf J}}
\newcommand{\bs}{{\bf v}_s}
\newcommand{\bn}{{\bf v}_n}
\newcommand{\bv}{{\bf v}}
\newcommand{\la}{\langle}
\newcommand{\ra}{\rangle}
\newcommand{\dg}{\dagger}
\newcommand{\br}{{\bf{r}}}
\newcommand{\brp}{{\bf{r}^\prime}}
\newcommand{\bq}{{\bf{q}}}
\newcommand{\hx}{\hat{\bf x}}
\newcommand{\hy}{\hat{\bf y}}
\newcommand{\bS}{{\bf S}}
\newcommand{\cU}{{\cal U}}
\newcommand{\cD}{{\cal D}}
\newcommand{\bR}{{\bf R}}
\newcommand{\pll}{\parallel}
\newcommand{\sumr}{\sum_{\vr}}
\newcommand{\cP}{{\cal P}}
\newcommand{\cQ}{{\cal Q}}
\newcommand{\cS}{{\cal S}}
\newcommand{\ua}{\uparrow}
\newcommand{\da}{\downarrow}
\newcommand{\red}{\textcolor {red}}
\newcommand{\blu}{\textcolor {blue}}
\newcommand{\1}{{\oldstylenums{1}}}
\newcommand{\2}{{\oldstylenums{2}}}
\newcommand{\mDelta}{\varepsilon}
\newcommand{\m}{\tilde m}
\def\lsim {\protect \raisebox{-0.75ex}[-1.5ex]{$\;\stackrel{<}{\sim}\;$}}
\def\gsim {\protect \raisebox{-0.75ex}[-1.5ex]{$\;\stackrel{>}{\sim}\;$}}
\def\lsimeq {\protect \raisebox{-0.75ex}[-1.5ex]{$\;\stackrel{<}{\simeq}\;$}}
\def\gsimeq {\protect \raisebox{-0.75ex}[-1.5ex]{$\;\stackrel{>}{\simeq}\;$}}

\title{Spatial correlations, additivity and fluctuations in conserved-mass transport processes}

\author{Arghya Das, Sayani Chatterjee and Punyabrata Pradhan} 

\affiliation{Department of Theoretical Sciences, S. N. Bose National Centre for Basic Sciences, Block-JD, Sector-III, Salt Lake, Kolkata 700098, India}

\begin{abstract}

\noindent{ We exactly calculate two-point spatial correlation functions in steady state in a broad class of conserved-mass transport processes, which are governed by chipping, diffusion and coalescence of masses.
We find that the spatial correlations are in general short-ranged and consequently, on a large scale, these transport processes possess a remarkable thermodynamic structure in the steady state. That is, the processes have an equilibriumlike additivity property and, consequently, a fluctuation-response relation, which help us to obtain subsystem mass distributions in the limit of subsystem size large.    
}

\typeout{polish abstract}

\end{abstract}


\maketitle

\section{Introduction}

Characterizing spatial structure in interacting many-particle systems having a nonequilibrium steady state (NESS) is a fundamental problem \cite{MatrixProduct_Derrida_JPhysA, Correlation_Derrida, Liggett, Privman}, though a difficult one, in statistical physics. In fact, the difficulty arises primarily because the exact steady-state weights of microscopic configurations, in most cases, are not known. A simple characterization, if any, of NESSs is highly desirable, especially when there is indeed a wide range of such systems in nature and also because they are arguably the closest counterpart to those in equilibrium. Like in equilibrium, the fluctuations in a NESS is time-stationary. However, unlike in equilibrium, these systems have nonzero currents and usually cannot be described by the Boltzmann distributions. Throughout the past couple of decades, there have been intensive studies to construct a suitable statistical mechanics framework having a unified description of systems having a NESS \cite{Gallavotti_Cohen_PRL, Oono, Lebowitz_Spohn_JSP1999, Bertini_PRL, Hatano-Sasa, Sasa_JSP, Sasa_PRE, Derrida_additivity}. Recently, a particular formulation, based on additivity, has emerged as a possible framework \cite{Eyink, Bertin, Pradhan_PRL2010}, which could not only help to have a unified characterization of a broad class of nonequilibrium systems, but could also be used to actually calculate density fluctuations in the systems \cite{Chatterjee_PRL2014, Chatterjee_PRE2015, Das_PRE2015}.

In this paper, we study a broad class of one dimensional conserved-mass transport processes involving chipping, diffusion and coalescence of masses and demonstrate that the processes possess, quite remarkably, an equilibriumlike thermodynamic structure. These mass transport processes have been studied intensively in the last couple of decades and have become a paradigm in nonequilibrium statistical physics of interacting many-particle systems. They represent a huge variety of natural processes, spanning a wide range of length scales, such as, formation of clouds \cite{cloud}, river networks \cite{river}, gels \cite{gel} and planets \cite{planet}, formation of lipid droplets on cell surface \cite{lipid}, fragmentation and self-assembly in various materials \cite{Vledouts_RSPA2016}, condensation of fluids on cold substrates \cite{condensation-fluid}, traffic flow \cite{traffic}, wealth distribution \cite{wealth} and migration and formation of cities \cite{migration}, etc.

The conserved-mass transport processes were first introduced as the Hammersley process \cite{Aldous} and as a model of force fluctuations in a pack of granular beads \cite{Majumdar_Science1995, Majumdar_PRE1996}. They were consequently generalized to various stochastic processes, called random average processes (RAPs) \cite{Ferrari, Rajesh_PRE2000, Krug_JSP2000, Zielen_JSP2002, Zielen_JPA2003, Kundu_condmat2015} or, equivalently, called mass chipping models (MCMs) \cite{Rajesh_JSP2000, Chatterjee_PRL2014, Mohanty_JSTAT2012}. There are also several other variants of these mass transport processes, which we call here mass exchange models (MEMs), where neighboring sites across a bond exchange among themselves a random fraction of their added masses \cite{wealth, CCModel}.

Though dynamical rules governing these processes are quite simple, they can give rise to nontrivial spatial structure in the steady state. In fact, even in one dimension which we consider in this paper, the exact steady-state weights, except for a few cases \cite{Krug_JSP2000, Rajesh_JSP2000, Zielen_JSP2002, Zielen_JPA2003, Mohanty_JSTAT2012, Kundu_condmat2015}, are not known. Notwithstanding the difficulty in obtaining the exact steady-state weights, there have been some progress in the past in calculating the two-point correlations in a few specific model systems \cite{Rajesh_PRE2000, Krug_JSP2000, Rajesh_JSP2000, Rajesh2_PRE2001, Kundu_condmat2015}. However, the spatial correlations for generic parameter values are still mostly unexplored. Moreover, another important quantity in these processes, the subsystem mass distribution when the subsystem size is large, or equivalently the large deviation function for subsystem mass, has not been studied when there are finite spatial correlations in the systems; single-site mass distributions have been calculated in the past, though for systems having a product-measure steady state \cite{Krug_JSP2000, Zielen_JSP2002} or within mean field theory \cite{Majumdar_PRE1996, Rajesh_JSP2000, Mohanty_JSTAT2012}.

In this paper, we characterize the steady-state spatial structure of these conserved-mass transport processes, by exactly calculating the two-point spatial (equal-time) correlations between masses at any two sites. Moreover, using an additivity property and a corresponding fluctuation-response (FR) relation, we demonstrate that, in the thermodynamic limit, the knowledge of only the two-point correlation functions is sufficient for obtaining the probability distribution function of mass in a subsystem, which is much larger than the spatial correlation length in the system. In other words, in the conserved-mass transport processes, we provide a formulation to obtain the large deviation probability of subsystem masses. Analogous to equilibrium free energy, the logarithm of the large deviation probability can be considered as a nonequilibrium free energy function, which governs the density fluctuations in these nonequilibrium processes and thus immediately connects to the standard statistical mechanics framework.

The organization of the paper is as follows. In section II. we discuss additivity and how the additivity can be used to obtain subsystem mass distributions in nonequilibrium systems. In section III, we exactly calculate the two-point spatial correlations in the three variants of the mass chipping models (MCMs) - mass chipping models I (MCM I) in section III.A, mass chipping models II (MCM II) in section III.B and mass chipping models III (MCM III) in Sec. III.C.  In section IV, we calculate two-point spatial correlations in mass exchange model (MEM) and then we summarize with some concluding remarks.

\section{Additivity and subsystem mass distribution}

In this section, we explain how additivity property can be used to calculate subsystem mass distribution when the subsystem size is large, irrespective of whether the system is in or out of equilibrium.

Let us discuss additivity first in the context of equilibrium. For equilibrium systems having an energy function $E$ with short-range interactions, the microscopic weight of a configuration $C$ can be written in terms of the Boltzmann distribution $P(C) \sim \exp[-\beta E(C)]$ where $\beta$ is the inverse temperature. It is well known that, in the thermodynamic limit, such an equilibrium system can be divided into many large subsystems which, being large and thus statistically almost independent, can be characterized using thermodynamic potentials like entropy or free energy function. For example, joint distribution $P[\{N_k\}]$ of subsystem particle-numbers $\{N_1, N_2, \dots, N_\nu \}$ in a system of volume $V$, which is kept in contact with a heat bath of inverse temperature $\beta$ and has a fixed total particle-number $N$, can be obtained from free energy function $F(N_k, v)$ of the individual subsystems of volume $v$,
\be
P[\{N_k\}] \simeq \frac{\prod_{k=1}^{\nu} e^{-\beta F(N_k, v)}}{e^{-\beta F(N, V)}} \delta \left( \sum_{k=1}^{\nu} N_k - N \right),
\ee   
where $N_k$ is the number of particles in the $i$th subsystem. The free energy function $F(N, V) = - \ln \{\sum_C \exp[-\beta E(C)] \}$ can in principle be calculated from the Boltzmann weights. The property that the joint subsystem distribution $P[\{ N_k \}]$ for a system can be approximately written as a product (i.e., subsystems are almost independent) of individual subsystem weight factors $\exp[-\beta F(N_k)]$ is called additivity property, which remains to be the corner-stone in equilibrium thermodynamics.

However, for systems having a NESS, there is usually no internal energy function, which can lead to the microscopic probability weights of the steady-state configurations, nor there is any well-defined notion of thermodynamic potentials as in equilibrium. In fact, for most of these nonequilibrium systems, the steady-state weights are not a-priori known and, to find them, one usually requires to explicitly obtain the time-independent solution of the Master equation (here we consider only the systems, which are governed by stochastic Markovian dynamics). Precisely at this stage, the difficulty arises as, in a driven many-particle system, it is often a formidable task to find these detailed microscopic weights. However, as demonstrated recently in Refs. \cite{Chatterjee_PRL2014, Das_PRE2015}, to characterize the fluctuation properties of a macroscopic quantity, such as the distribution of mass in a large subsystem, one may not actually need to calculate the weights of all microscopic configurations; rather, obtaining coarse-grained probability weights on a larger scale would suffice for this purpose, provided additivity, as discussed below (see Eq. \ref{Additivity}), holds.

As illustrated in this paper, for obtaining the large-scale fluctuation properties of a system, which could be in or out of equilibrium, what we need to know a-priori is an additivity property: Large subsystems should be statistically almost independent. Additivity is physically quite expected provided that the subsystems are much larger than the spatial correlation length in the system so that the boundary correlations between the subsystems could be ignored. In other words, the joint probability  distribution of subsystem masses $\{M_1, M_2, \dots, M_{\nu}\}$, to a good approximation, can be written in a product form,
\be
P[\{M_k\}] \simeq \frac{\prod_{i=1}^{\nu} W_v(M_k)}{Z(M,V)}\delta \left(\sum_{k=1}^{\nu} M_k - M \right),
\label{Additivity}
\ee
where the weight factor $W_v(M_k)$, still unknown and to be determined later, is assumed to depend only on the respective subsystem mass $M_k$, $v$ is the volume of each subsystem and $\nu$ is total number of subsystems. In the above equation, the normalization constant, or the partition sum, can be written as
$$
Z(M,V) = \left[ \prod_{k} \int dM_k W_v(M_k) \right] \delta \left( \sum_k M_k - M \right).
$$ 
Now, the probability that mass $M_k$, say, in the $k$th subsystem, lies in the interval $(m,m+dm)$ can be expressed as ${\rm Prob}[M_k \in (m,dm)] \equiv P_v(m) dm$ where the probability density function can be formally written, in the limit of large subsystem size $v$, as
\be
P_v(m) \simeq \frac{1}{\cal Z} W_v(m) e^{\mu(\rho) m}
\label{Pv_m}
\ee
with $\mu(\rho)$ a chemical potential, $\rho=M/V$ mass density and
\be
{\cal Z}(\mu) = \int_0^{\infty} W_v(m) e^{\mu m} dm
\label{Z_mu}
\ee 
the normalization constant. Moreover, the weight factor and chemical potential can be determined from a nonequilibrium 
fluctuation-response (FR) relation between nonequilibrium compressibility $d\rho/d\mu$ and the subsystem particle-number fluctuation, 
\be
v \frac{d\rho}{d\mu} = \sigma_v^2(\rho)
\label{FR},
\ee
which, as discussed below, is a direct consequence of additivity \cite{Eyink, Bertin, Pradhan_PRL2010, Chatterjee_PRL2014}; here, $\sigma_v^2(\rho) = \langle M_k^2 \rangle - \langle M_k \rangle^2$ is the variance, or the standard deviation, of mass in the $k$th subsystem and is a function of density $\rho$. 
The above nonequilibrium FR relation has indeed a very close resemblance with the familiar fluctuation dissipation theorem (FDT) in equilibrium, where compressibility is related to particle number fluctuation in a system. At this stage, it is not difficult to see why the quantity $\mu(\rho)$ can be interpreted as an equilibriumlike chemical potential even for systems having a NESS. In fact, the FR relation can be proved using Eqs. \ref{Pv_m} and \ref{Z_mu}; for the sake of completeness, the proof is provided below. We first note that the mean and the variance of subsystem mass $\langle M_k \rangle$ and $\langle M_k^2 \rangle - \langle M_k \rangle^2$, respectively, can be written as 
\bea
\langle M_k \rangle = v \rho = \frac{d \ln {\cal Z}}{d \mu}, \label{m_avg}
\\
\langle M_k^2 \rangle - \langle M_k \rangle^2 = \frac{d^2 \ln {\cal Z}}{d \mu^2}. \label{m_var}
\eea
Now taking derivative of Eq. \ref{m_avg} w.r.t. chemical potential and then using  Eq. \ref{m_var}, we obtain the FR as in Eq. \ref{FR}.

As illustrated later in various model systems, the variance of subsystem mass as a function of density can be obtained from the knowledge of two-point spatial correlations of microscopic mass variables at two lattice sites. Once we obtain the functional dependence of the variance $\sigma_v^2(\rho)$ on mass density $\rho$, we immediately have expressions for chemical potential, 
\be
\mu(\rho) = \int \frac{1}{\sigma^2(\rho)} d \rho + \alpha,
\label{mu}
\ee 
where we define $\sigma^2(\rho)=\sigma_v^2(\rho)/v$ the scaled variance in the thermodynamic limit and, consequently, a free energy density function, 
\be
f(\rho)=\int \mu(\rho) d\rho + \alpha \rho + \beta,
\label{free-energy}
\ee 
by twice integrating the FR relation w.r.t. density, where $\alpha$ and $\beta$ are arbitrary constants of integration \cite{Chatterjee_PRL2014}. Then, Laplace transform of the weight factor,
\be
\tilde{W}_v(s) = \int_0^{\infty} W_v(m) e^{-s m} dm \equiv e^{-\Lambda_v(s)},
\label{LT}
\ee 
can be obtained from the function $\Lambda_v(s)$ using Legender transform of free energy density \cite{Das_PRE2015, Touchette},
\be
\Lambda_v(s) = v [\mbox{\bf inf}_{\rho} \{ f(\rho) + s \rho \} ] = v[f(\rho^*) + s \rho^*],
\label{ls1}
\ee
where $\rho^*(s)$ is the solution of $s=-f'(\rho^*)$, i.e., 
\be
 s=-\mu(\rho^*).
\label{rho-star}
\ee 
Now, performing inverse Laplace transform of $\tilde w_v(s)$, we get the weight factor $W_v(m)$ and, thereafter, substituting $\mu(\rho)$ obtained from Eq. \ref{mu} in Eq. \ref{Pv_m}, we obtain the probability density function $P_v(m)$ for subsystem mass.

In the subsequent sections, we calculate the variance $\sigma_v^2(\rho)$ of mass in a subsystem of size $v$ as a function of mass density $\rho$ in a broad class of conserved-mass transport processes. Interestingly, in all these cases, we find that the variance $\sigma_v^2(\rho)$ of subsystem mass has the following functional dependence on mass density $\rho$,
\be
\sigma_v^2(\rho) = v \frac{\rho^2}{\eta},
\ee  
i.e., the variance of subsystem mass is proportional to the square of mass density, where the factor $\eta$ depends on microscopic parameters of the particular model systems. In that case, chemical potential and free energy density can be immediately obtained from Eqs. \ref{mu} and \ref{free-energy},
\bea
\mu(\rho) = -\frac{\eta}{\rho} + \alpha,
\\
f(\rho) = - \eta \ln \rho + \alpha \rho + \beta,
\eea
which, using Eqs. \ref{LT}, \ref{ls1} and \ref{rho-star}, respectively, lead to the following expressions, 
\bea
s = \frac{\eta}{\rho^*} - \alpha,
\\
\Lambda_v(s) = {\rm const.} + \ln \left[ (s+\alpha)^{\eta v} \right],
\\
W_v(s) = {\rm const.} (s+\alpha)^{-v \eta}.
\eea
Now, performing inverse Laplace transform of $W_v(s)$, we obtain the weight factor,
\be
W_v(m) = {\rm const.}  m^{v \eta -1} e^{-\alpha m},
\ee
and the corresponding probability distribution function for subsystem mass,
\be
P_v(m) \propto m^{v\eta -1} e^{- \eta m/\rho},
\label{gamma}
\ee
which has the form of gamma distribution. The above subsystem mass distribution can be immediately recast as given below,
\be
P_v(m) \simeq {\rm const.}  e^{-v h(m/v)},
\ee
in the form of a large deviation function, or a rate function \cite{Touchette}, $h(x) = -\eta \ln x - \mu x$.

\section{Mass Chipping Model (MCM)}

\begin{figure}
\begin{center}
\leavevmode
\includegraphics[width=7.5cm,angle=0]{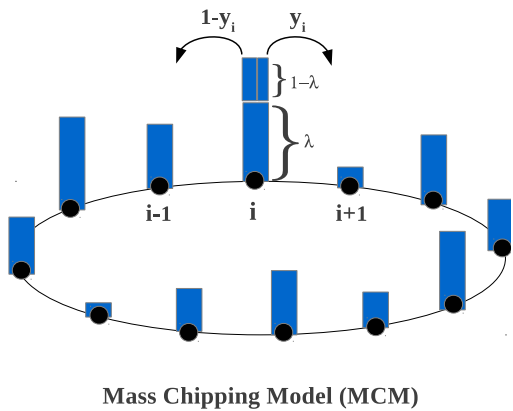}
\caption{Schematic representation of the mass chipping models (MCMs): $(1-\lambda)$ fraction of mass $m_i$ at a site $i$ is chipped off. Then, one or both of the fractions, $y_i$ and $(1-y_i)$, of the chipped-off mass diffuse and coalesce with one of the nearest-neighbor masses, depending on whether the mass transfer rule to the neighbors is asymmetric or symmetric. Random number $y_i \in [0,1]$ is drawn from a distribution $\phi(y_i)$. }
\label{FIG-MCM}
\end{center}
\end{figure}

In this section, we study mass chipping models (MCMs), which are defined on a one dimensional ring of $L$ sites, having a continuous mass variable $m_i \geq 0$ at site $i$ where total mass $M=\sum_{i=1}^L m_i$ remains conserved \cite{Ferrari, Rajesh_PRE2000, Krug_JSP2000, Rajesh_JSP2000, Zielen_JSP2002, Zielen_JPA2003, Rajesh2_PRE2001, Mohanty_JSTAT2012}. The dynamics involves chipping, diffusion and coalescence of masses. In the process of chipping, a site $i$ keeps a certain fraction $\lambda$ of its mass $m_i$ and the rest of the mass $(1-\lambda)m_i$ is chipped off. Then, a random fraction $y_i$ of the chipped-off mass, where $y_i$ is chosen from a probability density function $\phi(y_i)$ with $y_i \in [0,1]$, is transferred to one of its nearest neighbors. The rest of the chipped-off mass is either deposited back to the departure site or transferred to its other nearest neighbor. The mass-chipping processes are schematically represented in Fig. \ref{FIG-MCM}. Depending on the details of these dynamical rules, there can be several variants of the MCM, as discussed below.

In the first variant, which we call mass chipping model I (MCM I) - particular versions of which were studied in the context of asymmetric random average processes (ARAP) in Refs. \cite{Krug_JSP2000, Rajesh_JSP2000, Zielen_JSP2002, Zielen_JPA2003}, mass is transferred completely asymmetrically in a particular direction (say, clockwise) as following. A random fraction $y_i$ of the chipped-off mass, i.e., $y_i(1-\lambda)m_i$, is transferred {\it only} to the {\it right} nearest neighbor and the rest of the mass, i.e., $(1-y_i)(1-\lambda)m_i$, comes back to the departure site.  
In mass chipping model II (MCM II), which were introduced in Ref. \cite{Mohanty_JSTAT2012}, a random fraction $y_i$ of the chipped-off mass goes to the right neighbor and  the other fraction goes to the left neighbor. In mass chipping model III (MCM III), which is a generalized version of the models studied in Refs. \cite{Krug_JSP2000, Rajesh_JSP2000, Zielen_JSP2002}, the mass transfer rule is as follows. A random fraction $y_i$ of the 
chipped-off mass goes to {\it either} of the nearest neighbors, with equal probability $1/2$; the rest of the chipped-off mass is deposited back to the departure site.

Note that, in MCM I, mass is transferred completely asymmetrically to the right nearest neighbor; consequently, there is a mass-current in the system. On the contrary, in MCM III, mass is transferred completely symmetrically, with equal probability, to either of its nearest neighbors; in this case, there is {\it no} net current in the system. However, in MCM II, the mass transfer can be effectively either symmetric or asymmetric, depending on the form of the probability density function $\phi(y)$. For a symmetric probability density $\phi(y) = \phi(1-y)$, the mass transfer in MCM II is indeed symmetric (therefore, no net current in the system); otherwise, the mass transfer is effectively asymmetric and therefore there can be a net current in a particular direction.

Stochastic updates in all these variants of the MCMs are done according to either random sequential update (RSU) or parallel update (PU).

\subsection{Mass Chipping Model I (MCM I)}

\subsubsection{Random Sequential Update (RSU)}

In MCM I with random sequential update (RSU), a site $i$ is chosen randomly. A fraction $\tilde \lambda = (1-\lambda)$ of mass $m_i$ at site $i$ is chipped off and then a random fraction $y_i$ of this chipped-off mass $\tilde \lambda m_i$ is transferred to the right nearest neighbor; the rest, $\tilde \lambda (1-y_i) m_i$, of the chipped-off mass is deposited back to the site $i$. The dynamics in infinitesimal time $dt$ can be written as given below, 
\bea
m_i(t+dt) =
\left\{
\begin{array}{ll} 
 \mbox{\underline{value:}} & ~\mbox{\underline{probability:}} \cr
\lambda m_i(t)+ \tilde \lambda(1-y_i)m_i(t)  & ~dt, \cr
m_i(t)+ \tilde \lambda y_{i-1}m_{i-1}(t) & ~dt, \cr 
 m_i(t) &  ~(1-2 dt),
\end{array}
\right.
\label{MCMI}
\eea
where the mass value in the first column of r.h.s. is assigned to the mass $m_i(t+dt)$ at a particular site $i$ at time $t+dt$ with the corresponding probability given in the second column and $y_i \in [0,1]$ is a random variable having a probability density $\phi(y_i)$. The first two moments of the probability density function $\phi(y)$ are denoted as 
\bea
\theta_1 = \int_0^1 y \phi(y) dy,
\\ 
\theta_2 = \int_0^1 y^2 \phi(y) dy.
\eea 
For the purpose of demonstration, we choose throughout  in simulations a particular probability density $\phi(y) = 1$, i.e., a uniform distribution in the unit interval of $y \in [0,1]$, providing $\theta_1={1}/{2}$ and $\theta_2={1}/{3}$.

We now define two-point correlation function as $c_r=\mathcal{C}_r -\rho^2$ where $\mathcal{C}_r=\langle m_i m_{i+r}\rangle$ with $r \in \{0, 1, \dots, L-1 \}$. Note that, for $r=0$, the quantity $\mathcal{C}_0$ is actually the second moment of mass at any site. 
Using the update rules as in Eq. \ref{MCMI}, infinitesimal time-evolution of the first moment $\langle m_i(t) \rangle$, up to order $dt$, can be written as
\bea
\langle m^2_i(t+dt) \rangle = \langle m^2_i(t) \rangle (1-2dt)
\nonumber \\
+ \langle [\lambda+\tilde \lambda (1-y_i)]^2 m^2_i(t) \rangle dt \nonumber \\
+ \langle [m_i(t) + \tilde \lambda y_{i-1}m_{i-1}(t)]^2 \rangle dt + {\cal O}(dt^2),~~
\eea
or, equivalently,
\bea
\frac{d{\cal C}_0}{dt} = \frac{d \langle m_i^2(t) \rangle}{dt} = - 2 \langle m^2_i(t) \rangle  + \langle [\lambda+\tilde \lambda (1-y_i)]^2 m^2_i(t) \rangle \nonumber \\
+ \langle [m_i(t) + \tilde \lambda y_{i-1}m_{i-1}(t)]^2 \rangle. ~~~
\eea
By using the steady-state condition $d {\cal C}_0/dt =0$ and that the fact that $y_i$ and $m_i$ are independent random variables, we have
\be
\mathcal{C}_0 = \frac{\theta_1}{\theta_1-\theta_2(1-\lambda)}\mathcal{C}_1 
\label{amcm_n0}.
\ee

The time evolution of two-point correlations $C_r$, for $r=1$ and $r \ge 2$, in infinitesimal time $dt$ can be written as
\begin{widetext}
\bea
m_i m_{i+1} (t+dt) = 
\left
\{
\begin{array}{ll} 
\mbox{\underline{value:}} & ~~~~\mbox{\underline{prob.:}} \cr
[m_i(t)+\tilde \lambda y_{i-1}m_{i-1}(t)]m_{i+1}  & ~~~~dt, \cr
m_i[\lambda m_{i+1}(t)+\tilde \lambda (1-y_{i+1})m_{i+1}(t)] & ~~~~dt, \cr 
[m_{i+1}(t)+\tilde \lambda y_i m_i(t)] [\lambda m_i(t)+\tilde \lambda (1-y_i) m_i(t)] &  ~~~~dt, \cr
m_i(t) m_{i+1}(t) & ~~~~(1-3 dt),
\end{array}
\right. %
\label{AMCM-C1}
\\
m_im_{i+r}(t+dt) = 
\left
\{
\begin{array}{ll} 
\mbox{\underline{value:}} & ~~~~\mbox{\underline{prob.:}} \cr
[m_i(t)+\tilde \lambda y_{i-1} m_{i-1}(t)] m_{i+r}(t) & ~~~~dt, \cr
[\lambda+\tilde \lambda (1-y_i)] m_i(t) m_{i+r}(t) & ~~~~dt, \cr
[m_{i+r}(t)+\tilde \lambda y_{i+r-1} m_{i+r-1}(t)] m_i(t) & ~~~~dt, \cr
[\lambda+\tilde \lambda (1-y_{i+r})] m_i(t) m_{i+r}(t) & ~~~~dt, \cr
m_i(t) m_{i+r}(t) & ~~~~(1-4dt),
\end{array}
\right. 
\label{AMCM-Cn}
\eea
\end{widetext}
which, using the steady-state condition $d{\cal C}_r/dt=0$, lead to
\be
\mathcal{C}_2-2\mathcal{C}_1 + \frac{\theta_1 - \theta_2 (1-\lambda)}{\theta_1}\mathcal{C}_0 = 0 \label{amcm_n1},
\label{MCMI-C1}
\ee
for $r=1$ and
\be
\mathcal{C}_{r+1}-2\mathcal{C}_r + \mathcal{C}_{r-1} = 0,
\label{MCMI-Cr}
\ee
for $r \ge 2$. The above relations imply $\mathcal{C}_2 = \mathcal{C}_1$ and ${\cal C}_r = \rho^2$ for $r \ge 2$ and can be combined to finally obtain the following,
\begin{eqnarray}
c_r = \mathcal{C}_r - \rho^2 = 
\left \{
\begin{array}{ll}
\frac{\theta_2(1-\lambda)}{\theta_1-\theta_2(1-\lambda)} \rho^2 \mbox{~~~for $r=0$,}\\
0 \mbox{~~~~~~~~~~~~otherwise.}
\end{array}
\right.
\end{eqnarray}
It is important to note that the relations in Eqs. \ref{amcm_n0}, \ref{MCMI-C1} and \ref{MCMI-Cr} involve only two-point, not 
three-point or any higher order, correlations. This is because, in this process (as well as in the other processes considered later), the probability (or, equivalently, the transition rate) with which each mass-chipping event occurs in an infinitesimal time $dt$  depends neither on the mass of the departure site nor on that of the destination site (e.g., see the transition probabilities given in the respective column in Eqs. \ref{AMCM-C1} and \ref{AMCM-Cn}). This is true in general for any $n$-point correlations, i.e., the 
time-evolution of a particular $n$-point correlation involves only other $n$-point correlations, not $n+1$ or higher order correlations. In other words,  the Bogoliubov-Born-Green-Kirkwood-Yvon (BBGKY) hierarchy for the correlation functions closes for these mass transport processes (MCMs as well as MEM).

\begin{figure}
\begin{center}
\leavevmode
\includegraphics[width=8.5cm,angle=0]{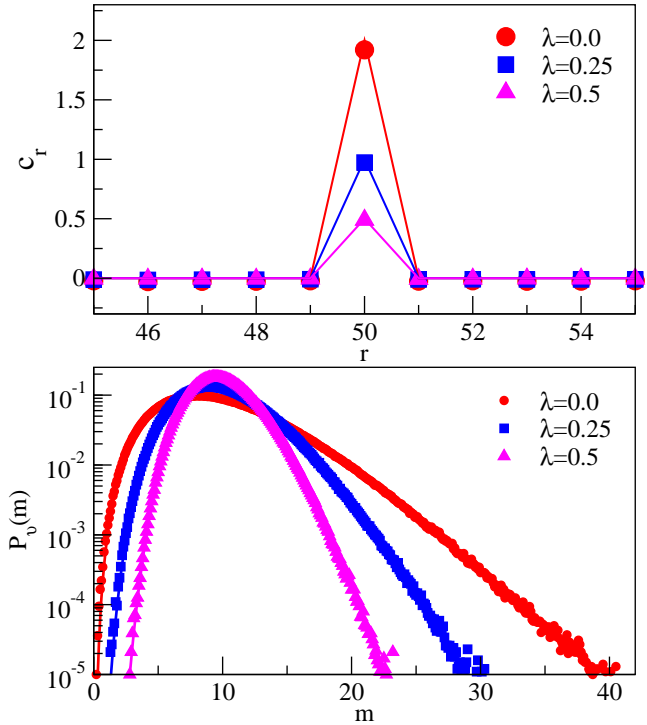}
\caption{Mass chipping model I (MCM I) with random sequential update (RSU) for $\lambda=0$, $0.25$ and $0.5$. In top panel, two-point correlation function $c_r = \langle m_i m_{i+r}\rangle - \rho^2$ is plotted as a function of distance $r$. In bottom panel, the probability density function $P_v(m)$ for mass in a subsystem size $v$ as a function of subsystem mass $m$. In all cases, system size $L=5000$, subsystem size $v=10$ and mass density $\rho=1$; points are simulations, lines are theory as in Eq. \ref{gamma}.  }
\label{AMCM-RSU}
\end{center}
\end{figure}

The variance of subsystem mass $m=\sum_{k=0}^{v-1} m_k$ can be written as $\sigma_v^2=\langle m^2\rangle - v^2 \rho^2$ where
\begin{equation}
\sigma_v^2 = v c_0 + 2(v-1) c_1 + 2(v-2) c_2 + \cdots + 2 c_{v-1}.
\label{sub_fluctuation}
\end{equation}
As $c_r = 0$ for $r \ne 0$, we obtain the variance of the subsystem mass, 
\begin{eqnarray}
\sigma_v^2=v c_0 = v \frac{\theta_2(1-\lambda)}{\theta_1-\theta_2 (1-\lambda)} \rho^2 \equiv v \frac{\rho^2}{\eta}
\end{eqnarray}
where
$$
\eta = \frac{\theta_1-\theta_2(1-\lambda)}{\theta_2(1-\lambda)}.
$$
Note that the variance is proportional to the square of the mass density. As derived in section II., this particular functional dependence of the variance on density, along with additivity Eq. \ref{Additivity}, implies that the subsystem mass distribution has the form of gamma distribution, 
\begin{equation}
P_v(m) = \frac{1}{\Gamma(v\eta)} \left(
\frac{\eta}{\rho} \right)^{v\eta} m^{v\eta - 1} 
e^{-\eta m/\rho}, \label{Pv} 
\end{equation} 
as in Eq. \ref{gamma} with the above expression of $\eta(\lambda, \theta_1, \theta_2)$. In Fig. \ref{AMCM-RSU}, we have compared our analytical results with the simulation results, where we numerically calculated the two-point correlation functions $c_r$ and the subsystem mass distributions $P_v(m)$ for various values of $\lambda=0$, $0.25$ and $0.5$ and for system size $L=5000$, $\rho=1$ and subsystem size $v=10$. Analytic and simulation results show very good agreement.

\subsubsection{Parallel Update (PU)}

\begin{figure}
\begin{center}
\leavevmode
\includegraphics[width=8.5cm,angle=0]{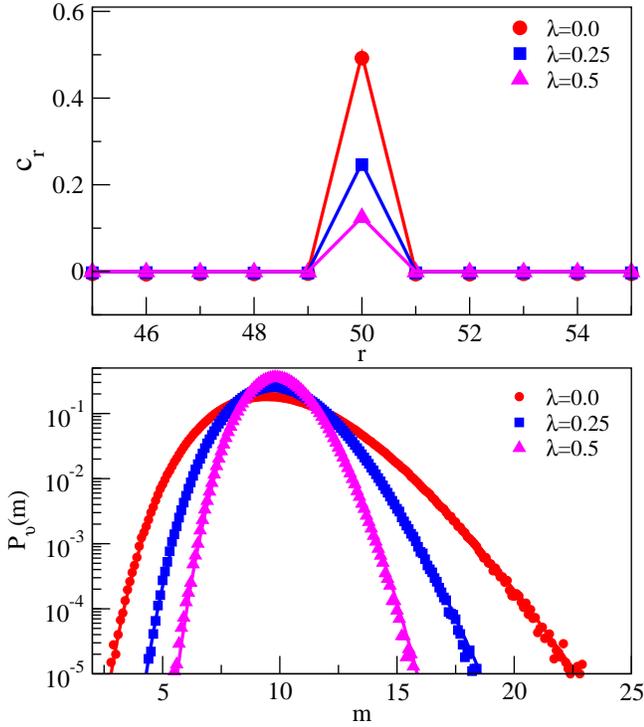}
\caption{Mass chipping model I (MCM I) with parallel update (PU) for $\lambda=0$, $0.25$ and $0.5$. In top panel, two-point correlation function $c_r=\langle m_i m_{i+r} \rangle - \rho^2$ is plotted as a function of distance $r$.
In bottom panel, the probability density function $P_v(m)$ for mass in a subsystem size $v$ as a function of subsystem mass $m$. In all cases, system size $L=5000$, subsystem size $v=10$, and mass density $\rho=1$; points are simulations, lines are theory as in Eq. \ref{gamma}.  }
\label{AMCM-PU}
\end{center}
\end{figure}

In MCM I with parallel update (PU), the amount of mass $\tilde \lambda y_i m_i(t)$, which is transferred from a site $i$ to the right nearest neighbor at a time step $t$, is the same as in the previous case of MCM I with RSU in Sec. III.A.1, but now all lattice sites are simultaneously updated in parallel. The discrete time stochastic dynamics with parallel update is given by,
\begin{equation}
m_i(t+1)=\lambda m_i(t)+ \tilde \lambda (1-y_i) m_i(t)+ \tilde \lambda y_{i-1} m_{i-1}(t)
\label{Dyn_AMCM_PU}
\end{equation}
where $y_i \in [0,1]$ is a random variable having probability density $\phi(y_i)$. The steady-state correlations can be 
calculated using the dynamical rules as in Eq. \ref{Dyn_AMCM_PU}.
We write below explicitly the discrete-time evolution of the two-point correlations $\langle m_i m_{i+r} \rangle$,
\begin{widetext}
\begin{eqnarray}
\langle m^2_i(t+1) \rangle = \lambda^2 \langle m^2_i(t) \rangle + \tilde \lambda^2  \langle (1-y_i)^2 m^2_i(t) \rangle
+ 2 \lambda \tilde \lambda \langle (1-y_i) m^2_i(t) \rangle 
+\tilde \lambda^2 \langle y_{i-1}^2 m^2_{i-1} \rangle 
~~~~~~~~\nonumber \\
+ 2 \langle [\tilde \lambda(\lambda + \tilde \lambda(1-y_i)) y_{i-1}] \langle m_i(t) m_{i-1}(t) \rangle 
\nonumber
\\
\langle m_i(t+1) m_{i+1}(t+1) \rangle = \langle [\lambda m_i(t)+ \tilde \lambda (1-y_i) m_i(t)+ \tilde \lambda y_{i-1} m_{i-1}(t)]  ~~~~~~~~~~~~~~~~~~~~~~~~~~~~~~~~~~~~~~~~~~\nonumber \\
\times [\lambda m_{i+1}(t)+ \tilde \lambda (1-y_{i+1})m_{i+1}(t)+ \tilde \lambda y_i m_i(t)] \rangle \nonumber
\\
\langle m_i (t+1) m_{i+r}(t+1) \rangle = \langle [\lambda m_i(t)+ \tilde \lambda (1-y_i) m_i(t)+ \tilde \lambda y_{i-1} m_{i-1}(t)]
~~~~~~~~~~~~~~~~~~~~~~~~~~~~~~~~~~~~~~~~~~ \nonumber \\
\times [\lambda m_{i+r}(t)+ \tilde \lambda (1-y_{i+r}) m_{i+r}(t)+ \tilde \lambda y_{i+r-1} m_{i+r-1}(t)] \rangle \nonumber
\end{eqnarray}
\end{widetext}
for $r=0$, $1$ and $r \ge 2$, respectively. Now using the steady-state condition $\langle m_i(t+1)m_{i+r}(t+1)\rangle = \langle m_i(t)m_{i+r}(t)\rangle$ in the above equations, we obtain the following relations between the correlation functions ${\cal C}_r$'s: For $r=0$,
\be
\mathcal{C}_0 = \frac{\lambda \theta_1+(1-\lambda)\theta_1(1-\theta_1)}{\lambda \theta_1+(1-\lambda)(\theta_1-\theta_2)}\mathcal{C}_1,
\label{MCMI-PU-0}
\ee
for $r=1$,
\begin{equation}
\mathcal{C}_2-2\mathcal{C}_1+\frac{\lambda \theta_1+(1-\lambda)(\theta_1-\theta_2)}{\lambda \theta_1+(1-\lambda)\theta_1(1-\theta_1)}\mathcal{C}_0=0
\label{MCMI-PU-1}
\end{equation}
and, for $r \geq 2$,
\begin{eqnarray}
\mathcal{C}_{r+1} -2\mathcal{C}_r + \mathcal{C}_{r-1} =0.
\label{MCMI-PU-r}
\end{eqnarray}
Note that the BBGKY hierarchy involving two-point correlation functions closes here. Now combining Eqs. \ref{MCMI-PU-0}, \ref{MCMI-PU-1}, and \ref{MCMI-PU-r}, we obtain the two-point correlation function,
%\textcolor{red}{
\begin{eqnarray}
c_r  = \mathcal{C}_r - \rho^2 =
\left \{
\begin{array}{ll}
\frac{(1-\lambda)(\theta_2-\theta_1^2)}{\lambda \theta_1+(1-\lambda)(\theta_1-\theta_2)} \rho^2 \mbox{~~for $r=0$}\\
0 \mbox{~~~~~~~~~~~~~~~~~~~~~~~\mbox{otherwise.}}
\end{array}
\right.
\end{eqnarray}
Consequently, we obtain the variance of subsystem mass, $\sigma_v^2 = vC_0 \equiv v {\rho^2}/{\eta}$ where 
\be
\eta=\frac{\lambda \theta_1+(1-\lambda)(\theta_1-\theta_2)}{(1-\lambda)(\theta_2-\theta_1^2)}. 
\ee
The subsystem mass distribution is given by gamma distribution as in Eq. \ref{gamma} with the above expression of $\eta(\lambda, \theta_1, \theta_2)$. 
In Fig. \ref{AMCM-PU}, we have compared our analytical and the simulation results for the two-point correlation function $c_r$ and the probability density function $P_v(m)$ for various values of $\lambda=0$, $0.25$ and $0.5$ and for system size $L=5000$, $\rho=1$ and subsystem size $v=10$. Analytic and simulation results show very good agreement.

Note that, for both random sequential and parallel update dynamics in MCM I, the two-point correlations vanish, $c_r=0$ for $r\ne 0$. In other words, spatial correlation length $\xi$ over which $c_r \sim \exp(-r/\xi)$ decays is essentially zero, i.e., $\xi=0$. In that case, additivity is expected to hold even on the single-site level, which is indeed the case as verified in \cite{Chatterjee_PRL2014} where the distribution of mass at any single site was shown to be well approximated by gamma distribution. In fact, as we have shown here, the variance calculated in \cite{Chatterjee_PRL2014} within mean field approximation is indeed exact as all the neighboring correlations vanish, i.e., $c_r=0$ for $r =1$.

\subsection{Mass Chipping Model II (MCM II)}

\subsubsection{Random Sequential Update (RSU)}

\begin{figure}
\begin{center}
\leavevmode
\includegraphics[width=8.5cm,angle=0]{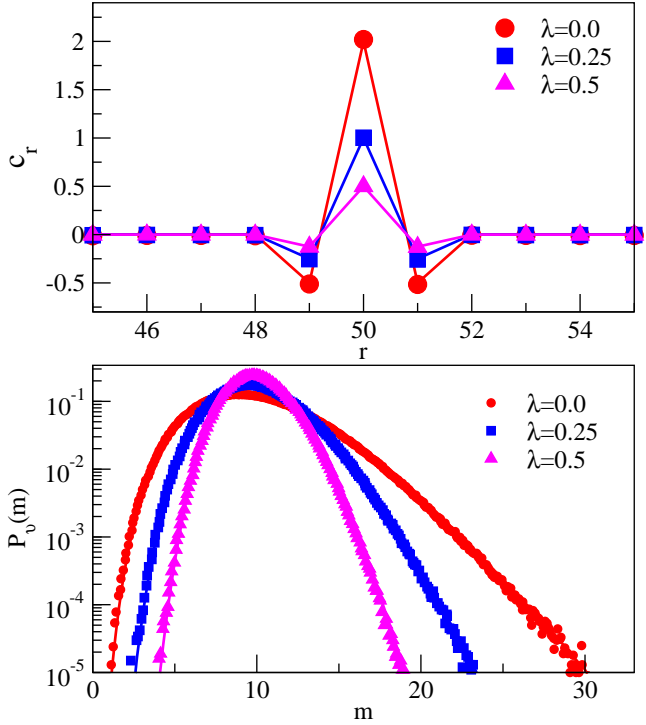}
\caption{Mass chipping model II (MCM II) with random sequential update (RSU) for $\lambda=0$, $0.25$ and $0.5$. In top panel, two point correlation function $c_r=\langle m_i m_{i+r} \rangle - \rho^2$ is plotted as a function of distance $r$.
In bottom panel, the probability density function $P_v(m)$ for mass in a subsystem size $v$ as a function of subsystem mass $m$. In all cases, system size $L=5000$, subsystem size $v=10$ and mass density $\rho=1$; points are simulations, lines are theory as in Eq. \ref{gamma}.   }
\label{SMCMI-RSU}
\end{center}
\end{figure}

In MCM II with random sequential update (RSU), a site $i$ is chosen randomly and a certain fraction $\tilde \lambda = 1-\lambda$ of mass $m_i$ at site $i$ is chipped off. Then, a random fraction $y_i$ of the chipped-off mass, i.e., $\tilde \lambda y_i m_i$ is transferred to the right nearest neighbor and the rest of the chipped-off mass, i.e., $\tilde \lambda (1-y_i) m_i$, is transferred to the left nearest neighbor \cite{Mohanty_JSTAT2012}.   
The stochastic update is given by, 
\begin{eqnarray}
m_i(t+dt) =  
\left\{
\begin{array}{ll} 
 \mbox{\underline{value:}} & \mbox{\underline{prob.:}} \cr
\lambda m_i(t) & dt \cr
m_i(t)+ \tilde \lambda y_{i-1} m_{i-1}(t) & dt \cr
m_i(t)+ \tilde \lambda (1-y_{i+1}) m_{i+1}(t) & dt \cr
m_i(t) & (1-3dt)
\end{array}
\right.
\end{eqnarray}
where $y_i \in [0,1]$ is a random variable having a probability density $\phi(y_i)$. Using the above time-evolution equation and the steady-state condition $d {{\cal C}_r}/dt = 0$, we get the following relations between two-point functions
\bea
\mathcal{C}_0 = \frac{1}{\lambda +(1-\lambda)(\theta_1-\theta_2)} \mathcal{C}_1 \label{smcm_n0},
\\
\mathcal{C}_2 - 2 \mathcal{C}_1 + \lambda \mathcal{C}_0 = 0 \label{smcm_n1},
\\
\mathcal{C}_3 -2 \mathcal{C}_2 + \mathcal{C}_1 + (1-\lambda)(\theta_1-\theta_2) \mathcal{C}_0 = 0 \label{smcm_n2},
\\
{\cal C}_{r+1} - 2 {\cal C}_r + {\cal C}_{r-1} = 0.
\eea
Solving the above equations, we obtain the two-point correlation function
\begin{eqnarray}
c_r = \mathcal{C}_r - \rho^2 = 
\left\{
\begin{array}{ll}
\frac{(1-\lambda)[1-2(\theta_1-\theta_2)]}{\lambda +2(1-\lambda)(\theta_1-\theta_2)} \rho^2 & \mbox{for $r=0$,} \cr
-\frac{(1-\lambda)(\theta_1-\theta_2)}{\lambda +2(1-\lambda)(\theta_1-\theta_2)}\rho^2 & \mbox{for $r=1$,} \cr
0 & \mbox{otherwise.}
\end{array}
\right.
\end{eqnarray}
Using Eq. \ref{sub_fluctuation}, we calculate the variance of subsystem mass $\sigma_v^2(\rho) =  v \rho^2/\eta$ as a function of density $\rho$ where
\begin{eqnarray}
\eta = \frac{\lambda +2(1-\lambda)(\theta_1-\theta_2)}{(1-\lambda)[1-2(\theta_1-\theta_2)(2-1/v)]}.
\end{eqnarray}
Therefore, the subsystem mass distribution is given by gamma distribution as in Eq. \ref{gamma} with the above expression of $\eta(\lambda, \theta_1, \theta_2)$. 
In Fig. \ref{SMCMI-RSU}, we have compared our analytical and the simulation results for the two-point correlation function $c_r$ and the probability density function $P_v(m)$ for various values of $\lambda=0$, $0.25$ and $0.5$ and for system size $L=5000$, $\rho=1$ and subsystem size $v=10$. Analytic and simulation results show very good agreement.

\subsubsection{Parallel Update (PU)}

In MCM II with parallel update (PU), the amount of masses which are transferred to the left and right are the same as in the case of RSU (see Sec. III.B.1), but now all sites are simultaneously updated in parallel \cite{Mohanty_JSTAT2012}. The update rule in this case is given below, 
\bea
m_i(t+1) = \lambda m_i(t)+ \tilde \lambda y_{i-1} m_{i-1}(t) + \tilde \lambda (1-y_{i+1}) m_{i+1}(t)~~~
\eea
where $\tilde \lambda = 1- \lambda$ and $y_i \in [0,1]$ is a random variable having probability density $\phi(y_i)$. In the steady state, the two-point correlations can be calculated using the above dynamics from the steady-state condition $\langle m_i(t+1)m_{i+r}(t+1)\rangle = \langle m_i(t)m_{i+r}(t)\rangle$, which gives the following relations between the ${\cal C}_r$'s: For $r=0$,
\begin{equation}
(\lambda+(1-\lambda)\beta)~\mathcal{C}_0- \lambda\mathcal{C}_1-(1-\lambda)\alpha~\mathcal{C}_2=0 \label{smcmP_n0},
\end{equation}
for $r=1$,
\begin{equation}
\lambda\mathcal{C}_0- (2\lambda+(1-\lambda)\alpha)~\mathcal{C}_1 +\lambda\mathcal{C}_2 +(1-\lambda)\alpha~\mathcal{C}_3=0 \label{smcmP_n1},
\end{equation}
and, for $r \geq 2$,
\bea
(1-\lambda)[\alpha +(\beta -\alpha)\delta_{r,2}] \mathcal{C}_{r-2} +\lambda\mathcal{C}_{r-1} \nonumber \\
-2[\lambda +(1-\lambda)\alpha] \mathcal{C}_r 
+\lambda \mathcal{C}_{r+1} +(1-\lambda)\alpha~\mathcal{C}_{r+2}=0 \label{smcmP_n}
\eea
where $\alpha = \theta_1(1-\theta_1)$ and $\beta = \theta_1-\theta_2$. To solve the above set of equations, we define a generating function,
\begin{equation}
G(z) = \sum_{r=1}^{\infty}  \mathcal{C}_r z^r,
\end{equation}
within a range of $|z| < 1$. Multiplying Eq. \ref{smcmP_n} by $z^r$ and summing over $r$, we obtain, using Eqs. \ref{smcmP_n0} and \ref{smcmP_n1},
\begin{equation}
G(z)=\frac{z}{(1-z)}\frac{z[\epsilon(1+z)+2(\kappa-1)]\mathcal{C}_0 +(1+z)\mathcal{C}_1}{(z-z_1)(z-z_2)}\label{genfn_primary}
\end{equation}
where
\bea
\epsilon = \frac{\beta}{\alpha}; ~ \kappa = 1+ \frac{\lambda}{2\alpha(1-\lambda)},
\\
z_1=-\kappa +\sqrt{\kappa^2 -1}; ~ z_2=\frac{1}{z_1} .
\eea
The quantities $\mathcal{C}_0$ and $\mathcal{C}_1$ can be obtained along the lines of arguments as in Ref. \cite{Rajesh_JSP2000}. Note that, in the limit of large $r$, $\mathcal{C}_r = \rho^2$ and, therefore, the asymptotic expression of the generating function is given by
\begin{equation}
\lim_{z\rightarrow 1}G(z)=\frac{\rho^2}{1-z},
\end{equation}
which, using Eq. \ref{genfn_primary}, immediately leads to a relation between $\mathcal{C}_0$ and $\mathcal{C}_1$
\begin{equation}
(\epsilon +\kappa -1)\mathcal{C}_0+\mathcal{C}_1= (1+\kappa)\rho^2.
\end{equation}
Moreover, at $z = z_1$, which is within the radius of convergence of generating $G(z)$, the function $G(z)$ appears to diverge. However, this cannot be the case as $G(z)$ must remain finite for $z_1 < 1$, implying that the numerator in the r.h.s. of Eq. \ref{genfn_primary} must vanish at $z_1$, leading to the second relation between $\mathcal{C}_0$ and $\mathcal{C}_1$,
\begin{equation}
z_1[\epsilon(1+z_1)+2(\kappa-1)]\mathcal{C}_0 +(1+z_1)\mathcal{C}_1=0\label{smcmpu_c0}.
\end{equation}
The last two equations give, 
\be
\mathcal{C}_0=\frac{1}{\epsilon +\sqrt{\frac{\kappa-1}{\kappa+1}}(1-\epsilon)}~\rho^2,
\ee and we obtain the variance of mass at a single site,
\begin{equation}
\sigma_1^2 = (1-\epsilon)\frac{1-\sqrt{\frac{\kappa-1}{\kappa+1}}}{\epsilon +\sqrt{\frac{\kappa-1}{\kappa+1}}(1-\epsilon)}\rho^2.
\end{equation}
Therefore, using Eq. \ref{smcmpu_c0} and the expression of $\mathcal{C}_0$ in (\ref{genfn_primary}), the final expression of the generating function is calculated to be,
\begin{equation}
G(z)=\epsilon~\mathcal{C}_0~\frac{z}{1-z}~\frac{\left(1+\frac{2\kappa -1}{\epsilon(1+z_1)}\right) +z}{z-z_2}
\end{equation}
After some algebraic manipulations, the two-point correlation function $c_r = \mathcal{C}_r - \rho^2$ can be expressed as
\begin{eqnarray}
c_r = 
\left\{
\begin{array}{ll}
(1-\epsilon)\frac{1-\sqrt{\frac{\kappa-1}{\kappa+1}}}{\left[\epsilon +\sqrt{\frac{\kappa-1}{\kappa+1}}(1-\epsilon)\right]}\rho^2 & \mbox{for $r=0$,} \cr 
-\frac{1}{1+\sqrt{\frac{\kappa+1}{\kappa-1}} \frac{\epsilon}{1-\epsilon}} z_1^r \rho^2 & \mbox{otherwise,}
\end{array}
\right.
\end{eqnarray}
where $z_1=-\kappa +\sqrt{\kappa^2 -1}$. The magnitude of the correlation function shows exponential decay $c_r \sim \exp(-r/\xi)$ where the correlation length $\xi$ is given by
\begin{equation}
\xi=-\frac{1}{\log{|z_1|}}.
\end{equation}

\begin{figure}
\begin{center}
\leavevmode
\includegraphics[width=8.5cm,angle=0]{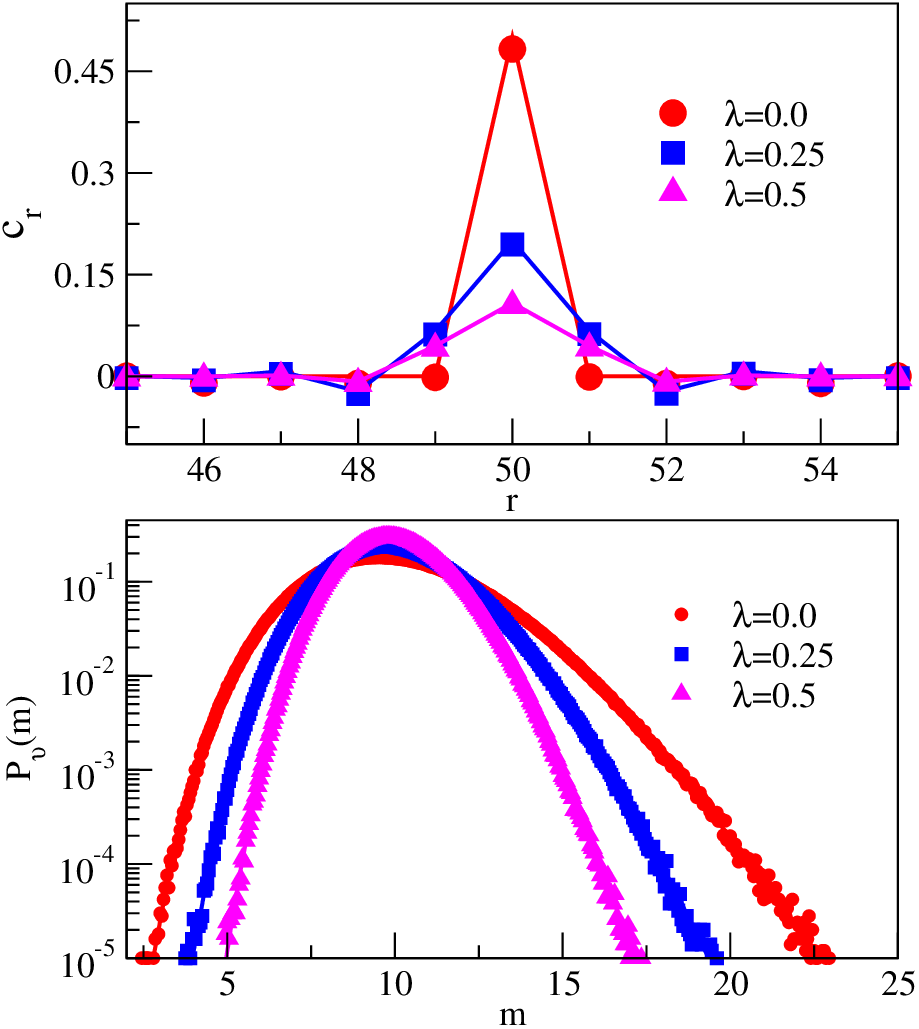}
\caption{Mass chipping model II (MCM II) with parallel update (PU) for $\lambda=0$, $0.25$ and $0.5$. In top panel, two-point correlation function $c_r=\langle m_i m_{i+r} \rangle - \rho^2$ is plotted as a function of distance $r$. In bottom panel, the probability density function $P_v(m)$ for mass in a subsystem size $v$ as a function of subsystem mass $m$. In all cases, system size $L=5000$, subsystem size $v=10$, and mass density $\rho=1$; points are simulations, lines are theory as in Eq. \ref{gamma}.   }
\label{SMCMI-PU}
\end{center}
\end{figure}

The variance of subsystem mass is obtained using Eq. \ref{sub_fluctuation},
\begin{eqnarray}
\sigma_v^2
= v \frac{2(1-\epsilon)}{\kappa +1} \frac{\left[1-\frac{1}{2v}\sqrt{\frac{\kappa -1}{\kappa +1}}~(1-z_1^v)\right]~\rho^2}{\left[\epsilon + \sqrt{\frac{\kappa -1}{\kappa +1}}(1-\epsilon)\right]} \equiv v \frac{\rho^2}{\eta}~~~~
\end{eqnarray}
where 
\be
\eta \approx \frac{\kappa+1}{2}~\left[\frac{\epsilon}{1-\epsilon} +\sqrt{\frac{\kappa -1}{\kappa +1}}\right]\left[1+\frac{1}{2v}\sqrt{\frac{\kappa -1}{\kappa +1}}\right].
\ee
Consequently, the subsystem mass distributions are described by gamma distribution as in Eq. \ref{gamma} with the above expression of $\eta(\lambda, \theta_1, \theta_2)$. In Fig. \ref{SMCMI-PU}, we have compared our analytical and the simulation results for the two-point correlation function $c_r$ and the probability density function $P_v(m)$ for various values of $\lambda=0$, $0.25$ and $0.5$ and for system size $L=5000$, $\rho=1$ and subsystem size $v=10$. Analytic and simulation results show very good agreement.

\subsection{Mass Chipping Model III (MCM III)}

\subsubsection{Random Sequential Update (RSU)}

In MCM III with random sequential update, a site $i$ is chosen randomly and a certain fraction $\tilde \lambda = \lambda$ of mass $m_i$ at site $i$ is chipped off. Further, a random fraction $y_i$ of the chipped-off mass, i.e., $\tilde \lambda y_i m_i$, is transferred either to the left or to right with equal probability $1/2$ and the rest of the chipped-off mass, i.e., $\tilde \lambda (1-y_i) m_i$, is deposited back to the site $i$. 
The stochastic time evolution in infinitesimal time $dt$ is given below, 
\begin{eqnarray}
m_i(t+dt) = 
\left\{
\begin{array}{ll} 
\mbox{\underline{value:}} & \mbox{\underline{prob.:}} \cr
\lambda m_i(t)+ \tilde \lambda (1-y_i) m_i(t) & dt \cr
m_i(t)+ \tilde \lambda y_{i+1} m_{i+1}(t) & {dt}/{2} \cr
m_i(t)+ \tilde \lambda y_{i-1} m_{i-1}(t) & {dt}/{2} \cr
m_i(t) & (1-2dt)
\end{array}
\right.
\label{SMCM-II-RSU-update}
\end{eqnarray}
$y_i \in [0,1]$ is a random variable having probability density $\phi(y_i)$.
Therefore, by putting $d{\cal C}_r/dt=0$ in the steady state, we obtain the following relations: For $r=1$,
\begin{equation}
[\theta_1 -(1-\lambda)\theta_2] \mathcal{C}_0- 2\theta_1\mathcal{C}_1+\theta_1\mathcal{C}_2=0,
\end{equation}
and, for $r \ge 2$,
\begin{eqnarray}
\mathcal{C}_{r+1}-2\mathcal{C}_r + \mathcal{C}_{r-1}=0,
\end{eqnarray}
implying $\mathcal{C}_r = \rm{constant}$ for $r \ge 2$. Finally, using the steady-state condition $d{\cal C}_0/dt=0$ and Eq. \ref{SMCM-II-RSU-update}, we obtain
\begin{equation}
\mathcal{C}_0=\frac{\theta_1}{\theta_1 -(1-\lambda)\theta_2}\mathcal{C}_1.
\end{equation}
Combining the above relations, we finally have the two-point correlation function,
\begin{eqnarray}
c_r = \mathcal{C}_r - \rho^2 =
\left\{
\begin{array}{ll} 
\frac{\theta_2(1-\lambda)}{\theta_1-\theta_2(1-\lambda)} \rho^2 & \mbox{for $r=0$,} \cr
0 & \mbox{otherwise},
\end{array}
\right.
\end{eqnarray}
which is interestingly identical to the results obtained for asymmetric mass chipping model with random sequential update. Accordingly, the variance of mass in a subsystem of size $v$ is given by
\begin{eqnarray}
\sigma_v^2=v c_0 \equiv v \frac{\rho^2}{\eta},
\end{eqnarray}
where 
$$
\eta =\frac{\theta_1-\theta_2(1-\lambda)}{\theta_2(1-\lambda)}.
$$ 
Therefore, the subsystem mass distribution is given by gamma distribution as in Eq. \ref{gamma} with the above expression of $\eta(\lambda, \theta_1, \theta_2)$. In Fig. \ref{SMCMI-PU}, we have compared our analytical and the simulation results for the two-point correlation function $c_r$ and the probability density function $P_v(m)$ for various values of $\lambda=0$, $0.25$ and $0.5$ and for system size $L=5000$, $\rho=1$ and subsystem size $v=10$. Analytic and simulation results show excellent agreement.

\begin{figure}
\begin{center}
\leavevmode
\includegraphics[width=8.5cm,angle=0]{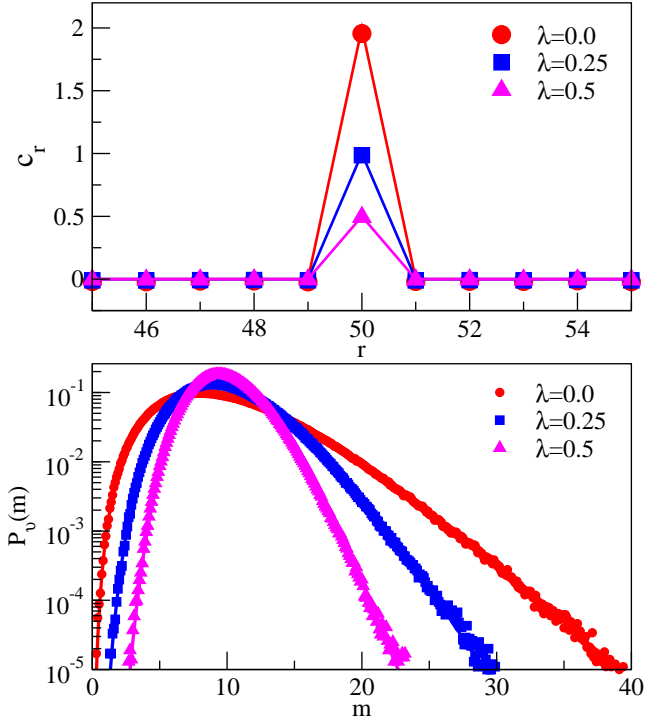}
\caption{Mass chipping model III (MCM III) with random sequential update (RSU) for $\lambda=0$, $0.25$ and $0.5$. In top panel, two-point correlation function $c_r=\langle m_i m_{i+r} \rangle - \rho^2$ is plotted as a function of distance $r$.
In bottom panel, the probability density function $P_v(m)$ for mass in a subsystem size $v$ as a function of subsystem mass $m$. In all cases, system size $L=5000$, subsystem size $v=10$, and mass density $\rho=1$; points are simulations, lines are theory as in Eq. \ref{gamma}.  }
\label{SMCMII-RSU}
\end{center}
\end{figure}

\subsubsection{Parallel update (PU)}

In MCM III with parallel update, the amount of mass which is transferred from a site $i$ is the same as in the case of RSU (see Sec. III.C.1), but now all sites are simultaneously updated in parallel. The discrete time-evolution is given below, 
\begin{eqnarray}
m_i(t+1) = [\lambda + \tilde \lambda (1-y_i)] m_i(t) + \tilde \lambda s_{i-1} y_{i-1} m_{i-1}(t)
\nonumber \\
 + \tilde \lambda (1-s_{i+1}) y_{i+1} m_{i+1}(t),~~
\end{eqnarray}
where we have introduced a random variable $s_i$ which takes discrete values $0$ and $1$, each with probability ${1}/{2}$. When the chipped-off mass moves to the right, $s_i = 1$ and otherwise $s_i=0$, implying $\langle s_i^n \rangle = {1}/{2}$ for $n \ne 0$. To calculate the two-point correlations, we use the steady-state condition $\langle m_i(t+1)m_{i+r}(t+1) \rangle = \langle m_i(t)m_{i+r}(t)\rangle$ to obtain, for $r=0$,
\begin{equation}
4[(1-\lambda)\epsilon -1] \mathcal{C}_0 +4\alpha\mathcal{C}_1 +(1-\alpha)\mathcal{C}_2=0,
\end{equation}
for $r=1$,
\begin{equation}
4[1-(1-\lambda)\epsilon] \mathcal{C}_0 -(1+7\alpha)\mathcal{C}_1 +4\alpha \mathcal{C}_2 +(1-\alpha)\mathcal{C}_3=0,
\end{equation}
and, for $r \ge 2$,
\bea
(1-\alpha)(1-\delta_{r,2})\mathcal{C}_{r-2} +4 \alpha \mathcal{C}_{r-1} -2(1+3 \alpha)\mathcal{C}_r 
\nonumber \\
+ 4 \alpha \mathcal{C}_{r+1} + (1-\alpha) \mathcal{C}_{r+2}=0,
\eea
where $\epsilon = {\theta_2}/{\theta_1}$ and $\alpha = 1-(1-\lambda)\theta_1$. As in the MCM II in Sec. III.B.2, one can readily solve these equations using the method of generating function $G(z) = \sum_{r=1}^{\infty} {\cal C}_r z^r$ as given below,
\begin{equation}
G(z)=\frac{1}{1-\alpha} \frac{z}{1-z} \frac{4[1-(1-\lambda)\epsilon] z\mathcal{C}_0 +(1-\alpha)(1+z)\mathcal{C}_1}{(z-z_1)(z-z_2)} \label{genfn_2}
\end{equation}
where 
\begin{equation}
z_1 =-\frac{1-\sqrt{\alpha}}{1+\sqrt{\alpha}}; ~~z_2 = \frac{1}{z_1}.
\end{equation}
Now, we obtain
\begin{equation}
2[1-(1-\lambda)\epsilon] \mathcal{C}_0 +(1-\alpha)\mathcal{C}_1=2\rho^2 
\end{equation}
as $\lim_{z \rightarrow 1}G(z)={\rho^2}/{(1-z)}$ and 
\begin{equation}
4 z_1 [1-(1-\lambda)\epsilon] \mathcal{C}_0 +(1-\alpha)(1+z_1)\mathcal{C}_1=0 \label{I_2}
\end{equation}
as the numerator of $G(z)$ is zero $z = z_1$. Eliminating $\mathcal{C}_1$ from the above two equations and using the expression of $z_1$, we obtain,
\begin{equation}
\mathcal{C}_0=\frac{\sqrt{\alpha}}{1-(1-\lambda)\epsilon}\rho^2
\end{equation}
The expression of $G(z)$ in Eq. \ref{genfn_2} then reduces to
\begin{equation}
G(z)=\frac{1}{1-\alpha} \frac{z}{1-z} \frac{4[1-(1-\lambda)\epsilon] \mathcal{C}_0 +(1-\alpha)\mathcal{C}_1}{z-z_2} \label{C0_2},
\end{equation}
which, from Eq. \ref{I_2}, is further reduced to
$$
G(z)=\frac{2}{1-\alpha}\frac{z}{1-z} \frac{[1-(1-\lambda)\epsilon] \mathcal{C}_0+\rho^2}{z-z_2}.
$$
Using $\mathcal{C}_0$ from Eq. \ref{C0_2}, we finally obtain
\begin{equation}
G(z)=\frac{z}{1-z}\frac{1-z_1}{1-zz_1}\rho^2,
\end{equation}
implying $\mathcal{C}_r =(1-z_1^r)\rho^2$ and therefore the two-point correlation function $c_r=\mathcal{C}_r - \rho^2$ can be written as
\begin{eqnarray}
c_r =
\left\{
\begin{array}{ll} 
\frac{(1-\lambda)\epsilon -(1-\sqrt{\alpha})}{1-(1-\lambda)\epsilon}\rho^2 & \text{~~for~} r=0, \cr
-z_1^r \rho^2 & \text{~~otherwise}.
\end{array}
\right.
\end{eqnarray}
Consequently, using Eq. \ref{sub_fluctuation}, the variance of subsystem mass is given by 
\begin{eqnarray}
{\sigma_v^2}(\rho) = v \left[\frac{(1-\lambda)\sqrt{\alpha}\epsilon}{1-(1-\lambda)\epsilon}-\frac{1-\alpha}{2v}(1-z_1^v)\right]\rho^2 \equiv v \frac{\rho^2}{\eta}
\end{eqnarray}
where
$$
\eta \approx \frac{1-(1-\lambda)\epsilon}{\sqrt{\alpha}(1-\lambda)\epsilon-\frac{1-\alpha}{2v}[1-(1-\lambda) \epsilon]}.
$$

\begin{figure}
\begin{center}
\leavevmode
\includegraphics[width=8.5cm,angle=0]{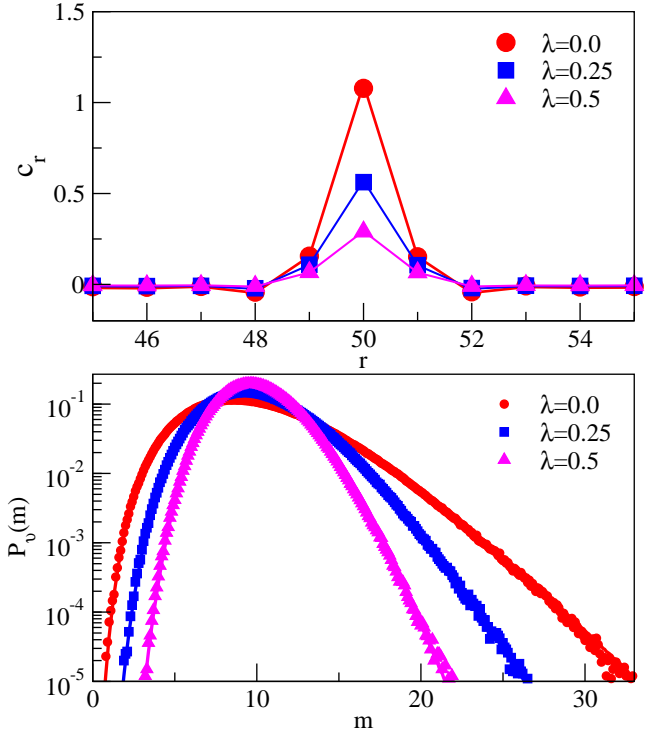}
\caption{Mass chipping model III (MCM III) with parallel update (PU) for $\lambda=0$, $0.25$ and $0.5$. In top panel, two-point correlation function $c_r=\langle m_i m_{i+r} \rangle - \rho^2$ is plotted as a function of distance $r$.
In bottom panel, the probability density function $P_v(m)$ for mass in a subsystem size $v$ as a function of subsystem mass $m$. In all cases, system size $L=5000$, subsystem size $v=10$, and mass density $\rho=1$; points are simulations, lines are theory as in Eq. \ref{gamma}.  }
\label{SMCMII-PU}
\end{center}
\end{figure}

Therefore, the subsystem mass distributions are described by gamma distribution as in Eq. \ref{gamma} with the above expression of $\eta(\lambda, \theta_1, \theta_2)$. 
In Fig. \ref{SMCMII-PU}, we have compared our analytical and the simulation results for the two-point correlation function $c_r$ and the probability density function $P_v(m)$ for various values of $\lambda=0$, $0.25$ and $0.5$ and for system size $L=5000$, $\rho=1$ and subsystem size $v=10$. Analytic and simulation results show excellent agreement.

\section{Mass exchange model (MEM)}

\begin{figure}
\begin{center}
\leavevmode
\includegraphics[width=7.5cm,angle=0]{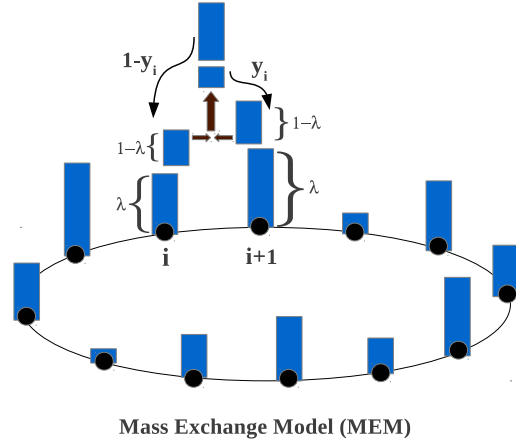}
\caption{Schematic representation of the mass exchange model (MEM): $(1-\lambda)$ fractions of masses $m_i$ and $m_{i+1}$ at two nearest neighbor sites $i$ and $i+1$ are chipped off and are added up. Then, $y_i$ and $(1-y_i)$ fractions of the added-up masses are assigned to one of the sites $i+1$ and $i$, respectively. Random number $y_i \in [0,1]$ is drawn from a distribution having a density function $\phi(y_i)$. }
\end{center}
\end{figure}

Mass exchange models have been studied throughout the past couple of decades \cite{Redner_EPJB, CCModel, Yakovenko, Patriarca_EPJB2010}, usually on a mean field level - on a graph where all sites interact with each other. Earlier, to consider the effect of a lattice structure on these processes, we studied the MEM on a one dimensional lattice \cite{Chatterjee_PRL2014}, where only neighboring masses can interact by exchanging certain fraction of masses among themselves. The lattice variant of the MEM gives rise to nontrivial spatial correlations, where the exact steady-state weights of the microscopic configurations, even in one dimension, are still unknown.

In this section, we exactly calculate the two-point spatial correlations for the MEM on a one dimensional periodic lattice of $L$ sites. The dynamical rules for the MEM are as follows. A bond between any two neighboring sites $i$ and $i+1$ is chosen randomly. Certain $\tilde \lambda = 1-\lambda$ fraction of their masses, i.e., $\tilde \lambda m_i$ and $\tilde \lambda m_{i+1}$, are chipped-off and added up. Then, $y_i$ and $1-y_i$ fractions of this added-up mass, where $y_i \in [0,1]$ is drawn from a distribution having a density function $\phi(y_i)$, are exchanged between the sites $i$ and $i+1$. Equivalently, the dynamical rules can be written as
\begin{eqnarray}
m_i(t+dt)= 
\left\{
\begin{array}{ll} 
\mbox{\underline{value:}} & \mbox{\underline{prob.:}} \cr
\lambda m_i(t)+ \tilde \lambda (1-y_i) m_{i,i+1}(t) & dt \cr
\lambda m_i(t)+ \tilde \lambda y_{i-1} m_{i-1,i}(t) & dt \cr
m_i(t) & (1-2dt)
\end{array}
\right.
\end{eqnarray}
where we define a bond-variable $m_{i,i+1}=m_i + m_{i+1}$ being total mass at the bond $(i,i+1)$. For the MEM, we consider only random sequential update as, in this case, parallel update is not well defined. Now using the above time-evolution equations and the steady-state condition $d{\cal C}_r/dt = 0$, we obtain the following relations. For $r=0$, the second moment of the distribution of mass at a single site is given by
\begin{equation}
\mathcal{C}_0=\frac{1-2(1-\lambda)(\theta_1 -\theta_2)}{\lambda +2(1-\lambda)(\theta_1 -\theta_2)}\mathcal{C}_1,
\end{equation} 
and, for $r=1$,
\begin{equation}
2\mathcal{C}_1=\mathcal{C}_2 + [\lambda +2(1-\lambda)(\theta_1 -\theta_2)] \mathcal{C}_0 +2(1-\lambda)(\theta_1 -\theta_2)\mathcal{C}_1.
\end{equation}
The above two relations imply, $\mathcal{C}_1 = \mathcal{C}_2$. Furthermore, for $r \ge 2$, we have 
\be
{\cal C}_{r+1} - 2 {\cal C}_r + {\cal C}_{r-1} = 0, 
\ee 
implying $\mathcal{C}_r = \rho^2$ for $r \ge 2$. Combining the all of the above relations, we finally obtain
\begin{eqnarray}
c_r = \mathcal{C}_r-\rho^2 = 
\left\{
\begin{array}{ll} 
\frac{(1-\lambda)[1-4(\theta_1 -\theta_2)]}{\lambda +2(1-\lambda)(\theta_1 -\theta_2)} \rho^2 & \text{for~~} r=0, \cr
0 & \text{otherwise,}
\end{array}
\right.
\end{eqnarray}
and consequently the variance of subsystem mass,
\begin{equation}
\sigma_v^2 = v c_0 \equiv v \frac{\rho^2}{\eta},
\end{equation}
where 
$$
\eta = \frac{\lambda +2(1-\lambda)(\theta_1 -\theta_2)}{(1-\lambda)[1-4(\theta_1 -\theta_2)]}.
$$ 
When the random number $y_i \in [0,1]$ is chosen from a uniform distribution $\phi(y_i)=1$, $\theta_1= 1/2$ and $\theta_2= 1/3$ and therefore $\eta(\lambda) = (1+2\lambda)/(1-\lambda)$. This particular expression of $\eta(\lambda)$ was obtained earlier within mean field approximation $\langle m_i m_{i+r} \rangle \approx \rho^2$ \cite{Chatterjee_PRL2014}, which, as we have shown in this section, is indeed exact due to the fact all the neighboring correlations vanish, i.e., $c_r = \langle m_i m_{i+r} \rangle - \rho^2 =0$ for $r \ge 1$. As demonstrated in the previous simulations \cite{Chatterjee_PRL2014}, the subsystem mass distributions for various values of $\lambda$ are indeed described by gamma distribution.

\section{Summary and concluding remarks}

In this paper, we have characterized spatial structure in a broad class of conserved-mass transport processes, which represent a wealth of natural phenomena concerning fragmentation, diffusion and coalescence of masses. 
Except for a few spacial cases which have a product measure \cite{Krug_JSP2000, Rajesh_JSP2000, Zielen_JSP2002, Mohanty_JSTAT2012, Kundu_condmat2015}, e.g., mass chipping model I with parallel update and with $\lambda=0$ \cite{Mohanty_JSTAT2012}, these processes in general give rise to nontrivial steady-state structure, which, in most of the cases, are  not exactly known \cite{Zielen_JPA2003}. 
%However, spatial correlations can still be calculated in these processes. 
Here, in the thermodynamic limit, we have exactly calculated the two-point spatial (equal-time) correlation functions, which are found to be short-ranged. Remarkably, these processes possess an equilibriumlike thermodynamic structure: They have an additivity property (Eq. \ref{Additivity}) and, consequently, there exists a fluctuation-response (FR) relation (Eq. \ref{FR}) between the compressibility and the fluctuations, analogous to the equilibrium fluctuation dissipation theorem (FDT). 
To substantiate our claims, we have used additivity and the corresponding FR relation to obtain, in the thermodynamic limit, the probability distribution function, including the large-deviation probability and the corresponding large deviation function, of subsystem mass.

In all the cases studied here, the variance of subsystem mass is proportional to square of the mass density. This particular functional dependence of the variance of subsystem mass on mass density, together with additivity and the FR relation, leads to the subsystem mass distributions having the form of gamma distribution. Quite interestingly, gammalike distributions have been observed in various experiments in the past \cite{Majumdar_Science1995, Vledouts_RSPA2016}, which could be understood in the light of the results of this paper.

We note that the main reason due to which the two-point spatial correlations can be calculated in these mass transport processes is that the Bogoliubov-Born-Green-Kirkwood-Yvon (BBGKY) hierarchy involving correlation functions here closes, e.g., the two-point correlations do not depend on the three-point (or higher order) correlations, etc. Nevertheless, the full characterization of three-point and higher order spatial correlations is still lacking and remains to be an interesting open issue, understanding of which could shed some light on the exact microscopic steady-state structure in these systems.

In summary, we have demonstrated that a broad class of conserved-mass transport processes have an equilibriumlike thermodynamic structure. That is, like in equilibrium, the fluctuations in these processes can be characterized in terms of thermodynamic potentials, such as a nonequilibrium free energy function and a chemical potential. Our results could be significant, considering that it is not often that, in driven interacting-particle systems, two-point correlations \cite{Correlation_Derrida, Rajesh2_PRE2001, Priyanka_PRE2014} and, especially, the mass distributions \cite{Derrida_Lebowitz, Majumdar_Sire} can be calculated exactly. 
From an overall perspective, our work leaves open the possibility of a unified thermodynamic framework for driven systems in general.

\section{Acknowledgement}

We thank Pradeep Mohanty for useful discussions. S.C. acknowledges financial support from the Council of Scientific and Industrial Research, India [Grant No. 09/575(0099)/2012-EMR-I]. P.P. acknowledges financial support from the Science and Engineering Research Board, India (Grant No. EMR/2014/000719).


\begin{thebibliography}{99}

\bibitem{MatrixProduct_Derrida_JPhysA} B. Derrida, E. Domany and D. Mukamel, J. Stat. Phys. {\bf 69}, 667 (1992); B. Derrida, M. R. Evans, V. Hakim, and V. Pasquier, J. Phys. A {\bf 26}, 1493 (1993).

\bibitem{Correlation_Derrida} B. Derrida and M. R. Evans, J. Phys. I France {\bf 3}, 311 (1993). 

\bibitem{Liggett} T. Liggett, {\it Interacting Particle Systems} (Springer-Verlag, New York, 1985).

\bibitem{Privman} V. Privman, {\it Nonequilibrium Statistical Mechanics in One
Dimension} (Cambridge University Press, Cambridge, 2005).

\bibitem{Gallavotti_Cohen_PRL} G. Gallavotti and E. G. D. Cohen, Phys. Rev. Lett. {\bf 74}, 2694 (1995). 

\bibitem{Oono} Y. Oono and M. Paniconi, Prog. Th. Phys. Supp. {\bf 130}, 29 (1998).

\bibitem{Lebowitz_Spohn_JSP1999} J. L. Lebowitz and H. Spohn, J. Stat. Phys. {\bf 95}, 333 (1999).

\bibitem{Bertini_PRL} L. Bertini, A. De Sole, D. Gabrielli, G. Jona-Lasinio, and C. Landim, Phys. Rev. Lett. {\bf 87}, 040601 (2001). J. Stat. Phys. {\bf 107}, 635 (2002). L. Bertini, A. De Sole, D. Gabrielli, G. Jona-Lasinio and C. Landim, Rev. Mod. Phys. {\bf 87}, 593 (2015).

\bibitem{Hatano-Sasa} T. Hatano and S. Sasa, Phys. Rev. Lett. {\bf 86}, 3463 (2001).

\bibitem{Sasa_JSP} S. Sasa and H. Tasaki, J. Stat. Phys. {\bf 125}, 125 (2006). 

\bibitem{Sasa_PRE} K. Hayashi and S. Sasa, Phys. Rev. E {\bf 68}, 035104 (2003). 

\bibitem{Derrida_additivity} T. Bodineau and B. Derrida, Phys. Rev. Lett. {\bf 92}, 180601 (2004).

\bibitem{Eyink} G. L. Eyink, J. L. Lebowitz, and H. Spohn, J. Stat. Phys. {\bf 83}, 385 (1996).

\bibitem{Bertin} E. Bertin, O. Dauchot, and M. Droz, Phys. Rev. Lett. {\bf 96}, 120601 (2006). 

\bibitem{Pradhan_PRL2010} P. Pradhan, C. P. Amann, and U. Seifert, Phys. Rev. Lett. {\bf 105}, 150601 (2010). 

\bibitem{Chatterjee_PRL2014} S. Chatterjee, P. Pradhan, and P. K. Mohanty,
Phys. Rev. Lett. {\bf 112}, 030601 (2014).

\bibitem{Chatterjee_PRE2015} S. Chatterjee, P. Pradhan, and P. K. Mohanty, Phys. Rev. E {\bf 91}, 062136 (2015).  

\bibitem{Das_PRE2015} A. Das, S. Chatterjee, P. Pradhan, and P. K. Mohanty,
Phys. Rev. E {\bf 92}, 052107 (2015).

\bibitem{cloud} S. K. Friedlander, Smoke, Dust, and Haze (Wiley Interscience, New York, 1977).

\bibitem{river} A. E. Scheidegger, Int. Assoc. Sci. Hydrol. Bull. {\bf 12}, 15 (1967).

\bibitem{gel} R. M. Ziff, J. Stat. Phys. {\bf 23}, 241 (1980).

\bibitem{planet} J. Blum and G. Wurm, Annu. Rev. Astro. Astrophys. {\bf 46}, 21 (2008).

\bibitem{lipid} A. Dutta and D. Sinha, Sci. Rep. {\bf 5}, Article number: 13915 (2015).

\bibitem{Vledouts_RSPA2016} A. Vledouts, N. Vandenberghe and E. Villermaux, Proc. R. Soc. A {\bf 471}, 20150678 (2015). A. Vledouts, N. Vandenberghe and E. Villermaux, Proc. R. Soc. A {\bf 471}, 20150679 (2016). 

\bibitem{condensation-fluid} P. Meakin, Rep. Prog. Phys. {\bf 55}, 157 (1992).

\bibitem{traffic} D. Chowdhury, L. Santen, and A. Schadschneider, Phys.
Rep. {\bf 329}, 199 (2000).

\bibitem{wealth} V. M. Yakovenko and J. B. Rosser, Rev. Mod. Phys. {\bf 81},
1703 (2009).

\bibitem{migration} J. Ke, Z. Lin, Y. Zheng, X. Chen, and W. Lu, Phys. Rev. Lett. {\bf 97}, 028301 (2006). 

\bibitem{Aldous} D. Aldous and P. Diaconis, Probab. Theor. Relat. Fields {\bf 103}, 199 (1995).

\bibitem{Majumdar_Science1995} C.-h. Liu, S. R. Nagel, D. A. Schecter, S. N.
Coppersmith, S. N. Majumdar, O. Narayan and T. A. Witten, Science {\bf 269},
513 (1995)

\bibitem{Majumdar_PRE1996} S. N. Coppersmith, C.-h. Liu, S. N. Majumdar, O. Narayan, and T. A. Witten, Phys. Rev. E {\bf 53}, 4673 (1996).

\bibitem{Ferrari} P. A. Ferrari and L. R. G. Fontes, El. J. Prob. {\bf 3}, Paper no. 6 (1998).

\bibitem{Rajesh_PRE2000} R. Rajesh and S. N. Majumdar, Phys. Rev. E {\bf 62}, 3186 (2000). 

\bibitem{Krug_JSP2000} J. Krug and J. Garcia, J. Stat. Phys. {\bf 99}, 31 (2000).

\bibitem{Rajesh_JSP2000} R. Rajesh and S. N. Majumdar, J. Stat. Phys. {\bf 99}, 943 (2000).

\bibitem{Zielen_JSP2002} F. Zielen and A. Schadschneider, J. Stat. Phys.
{\bf 106}, 173 (2002).

\bibitem{Zielen_JPA2003} F. Zielen and A. Schadschneider, J. Phys. A: Math.
Gen. {\bf 36}, 3709 (2003).

%\bibitem{Rajesh1_PRE2001} R. Rajesh and Satya N. Majumdar, Phys. Rev. E {\bf 63}, 036114 (2001).

\bibitem{Rajesh2_PRE2001} R. Rajesh and S. N. Majumdar, Phys. Rev. E {\bf 64}, 036103 (2001).

%\bibitem{Gupta_PRE2007} S. Gupta, S. N. Majumdar, C. Godreche, and M. Barma, Phys. Rev. E {\bf 76}, 021112 (2007).

\bibitem{Mohanty_JSTAT2012} S. Bondyopadhyay and P. K. Mohanty, J. Stat. Mech. {\bf P07019} (2012).

\bibitem{Touchette} H. Touchette, Phys. Rep. {\bf 478}, 1 (2009). 

%\bibitem{Gupta_PRE2011} S. Gupta, M. Barma, U. Basu, and P. K. Mohanty, Phys. Rev. E {\bf 84}, 041102 (2011).

\bibitem{Priyanka_PRE2014} Priyanka, A. Ayyer, and K. Jain, Phys. Rev. E {\bf 90}, 062104 (2014).

\bibitem{Kundu_condmat2015} J. Cividini, A. Kundu, S. N. Majumdar, D. Mukamel, J. Phys. A: Math. Theor. {\bf 49} 085002 (2016). 

\bibitem{Redner_EPJB} S. Ispolatov, P. L. Krapivsky, and S. Redner, Eur. Phys. J. B {\bf 2}, 267 (1998).

\bibitem{CCModel} C. Kipnis, C. Marchioro, and E. Presutti, J. Stat. Phys. {\bf 27}, 65 (1982); A. Chakraborti and B. K. Chakrabarti, Eur. Phys. J. B {\bf 17}, 167 (2000). 

\bibitem{Yakovenko}  V. M. Yakovenko and J. B. Rosser,  Rev. Mod. Phys.
{\bf 81}, 1703 (2009).

\bibitem{Patriarca_EPJB2010} M. Patriarca, E. Heinsalu, and A. Chakraborti,
Eur. Phys. J. B {\bf 73}, 145 (2010).

\bibitem{Majumdar_Sire} S. N. Majumdar and C. Sire, Phys. Rev. Lett. {\bf 71}, 3729 (19993). 

\bibitem{Derrida_Lebowitz} B. Derrida, J. L. Lebowitz, and E. R. Speer, Phys. Rev. Lett. {\bf 87}, 150601 (2001); B. Derrida, J. L. Lebowitz, and E. R. Speer, Phys. Rev. Lett. {\bf 89}, 030601 (2002).

\end{thebibliography}
\end{document}